\documentclass[fleqn,12pt]{wlscirep}
\usepackage[utf8]{inputenc}
\usepackage[T1]{fontenc}
\usepackage{lineno}
\usepackage{float}
\usepackage{ragged2e}
\usepackage{caption}

\title{Observation of tunable chiral spin textures with nonlinear optics }

\author[1,+]{Youqiang Huang}
\author[2,+]{Tiago V. C. Ant\~ao}
\author[2]{Adolfo O. Fumega}
\author[1]{Mikko Turunen}
\author[1]{Yi Zhang}
\author[3]{Hanlin Fang}
\author[1]{Nianze Shang}
\author[1]{Juan C. Arias-Mu\~noz}
\author[1]{Fedor Nigmatulin}
\author[4]{Hao Hong}
\author[5]{Andrew S. Kim}
\author[1]{Faisal Ahmed}
\author[5]{Hyunyong Choi}
\author[3]{Sanshui Xiao}
\author[4]{Kaihui Liu}
\author[2,*]{Jose L. Lado}
\author[1,*]{Zhipei Sun}

\affil[1]{QTF Centre of Excellence, Department of Electronics and Nanoengineering, Aalto University, Espoo, Finland}
\affil[2]{Department of Applied Physics, Aalto University, Espoo, Finland}
\affil[3]{Department of Electrical and Photonics Engineering, Technical University of Denmark, Konges Lyngby, Denmark}
\affil[4]{State Key Laboratory for Mesoscopic Physics, School of Physics, Peking University, Beijing, China}
\affil[5]{Department of Physics and Astronomy, Seoul National University, Seoul, 08826, Republic of Korea}

\affil[*]{zhipei.sun@aalto.fi, jose.lado@aalto.fi}
\affil[+]{These authors contributed equally to this work}

\begin{abstract}
Chiral spin textures, such as spin spirals and skyrmions, are key to advancing spintronics by enabling ultrathin, energy-efficient memory, and high-density data storage and processing. However, their realization remains hindered by the scarcity of suitable host materials and the formidable experimental challenges associated with the characterization of these intricate chiral magnetic states. Here, we report the observation of tunable chiral magnetic textures in van der Waals magnet CrPS$_4$ with nonlinear optics. These tunable textures exhibit strong chiral third-order nonlinear optical responses, driven by interlayer and intralayer spin couplings under varying magnetic fields and temperatures. These pronounced chiral nonlinear optical responses highlight the potency and high sensitivity of the nonlinear optical readout for probing non-collinear magnetic orders. Moreover, our findings position van der Waals magnets and their heterostructures as an exceptional platform for reconfigurable spin-photonics and spintronics, unifying optical, electrical, and magnetic properties through unique intralayer and interlayer spin coupling properties and effective spin interaction between photons and electrons.
\end{abstract}

\begin{document}

\flushbottom
\maketitle
%
%
\thispagestyle{empty}

\noindent 

\section*{Main}
Chiral spin textures, such as skyrmions and spin spirals, are of great interest in spintronics for their potential to enable compact, low-power memory and logic devices. A major bottleneck to their practical realization has been the lack of suitable material platforms with controllable and stable magnetic interactions. The emergence of two-dimensional (2D) van der Waals (vdW) magnets has addressed this challenge, offering highly tunable magnetic properties and seamless integration into layered heterostructures~\cite{Gibertini2019,Zhang2024,Blei2021,Zhang2022_optical}. Notable examples, including CrI$_3$, FePS$_3$, and Cr$_2$Ge$_2$Te$_6$, exhibit robust magnetic order down to the monolayer limit~\cite{Lee2016,Gong_2017,Huang_2017,Cai2019,OFumega2020_Crmagnets,Sun2021,Yao2023}, establishing 2D magnets as a powerful platform for studying low-dimensional magnetism. Indeed, these materials have enabled the experimental investigation of complex spin configurations and have broadened opportunities for spintronic and quantum information technologies~\cite{Fert_2013,Zhang_2023,Back_2020,Wang_2022_skyrmions}. Recent developments in twisted magnetic systems~\cite{Song2021,Yang_2024,Xie2023} and spin-spiral multiferroics~\cite{Song_2022,Amini_2024,Fumega_2022,Anto2024} further highlight the potential of 2D vdW magnets to host rich and reconfigurable chiral spin textures. Nevertheless, the discovery of vdW compounds capable of supporting such states remains limited, constraining the exploration of their interplay with other emergent phenomena and impeding a full understanding of their underlying mechanisms.

These challenges are further intensified by the limitations of current techniques for characterizing complex spin textures. While neutron diffraction~\cite{Li_2024} is effective for detecting magnetic orders in bulk materials, it is less suitable for ultrathin multilayers. Optical methods, including magnetic optical Kerr effect~\cite{Lan2020,MolinaSnchez2020,Hendriks2021,Kato2023} and reflectance magneto-circular dichroism~\cite{Suzuki2022,Ando1992,Burch2018}, are sensitive to out-of-plane magnetization but fall short in probing in-plane or chiral spin configurations~\cite{Song2021}. Nonlinear optical processes, especially third-harmonic generation (THG) circular dichroism, present a promising alternative, as shown in studies on chiral molecules~\cite{Harada2018} and metasurfaces~\cite{Gandolfi2021,Toftul2024,Kim_2020,Tonkaev2024,Koshelev2023}. Despite this transformative potential, the application of nonlinear optics to complex spin textures remains substantially unexplored~\cite{Wu_2024, Ono_2024}, underscoring the need to leverage these methods for a deeper understanding of the magnetic, spintronic, and chiral nonlinear optical properties of 2D materials.

Here, we observe complex chiral magnetic orders in CrPS$_4$ with nonlinear optics and propose a microscopic spin model describing their emergence due to competing intralayer and interlayer magnetic couplings. Using chiral THG, we demonstrate that these frustrated magnetic states, arising from nearest- and next-nearest-interlayer exchange interactions, can be finely tuned by temperature and magnetic field. Our measurements of chiral THG reveal these tunable spin spiral phases under varying magnetic field magnitudes, showcasing chiral nonlinear optics as a powerful tool for probing non-collinear spin configurations inaccessible to conventional methods. 
These findings underscore the potential of 2D magnets and their heterostructures, particularly CrPS$_4$, as versatile platforms for integrating optical, electrical, and magnetic chirality. The ability to merge chiral light-matter interactions with tunable chiral spin textures opens exciting avenues for advancing nonlinear reconfigurable spin-photonics, spintronics, and next-generation information storage and processing technologies.

\section*{Tunable chiral magnetic orders in CrPS$_4$}

\begin{figure}[htbp]
    \centering
    \includegraphics[width=0.9\textwidth]{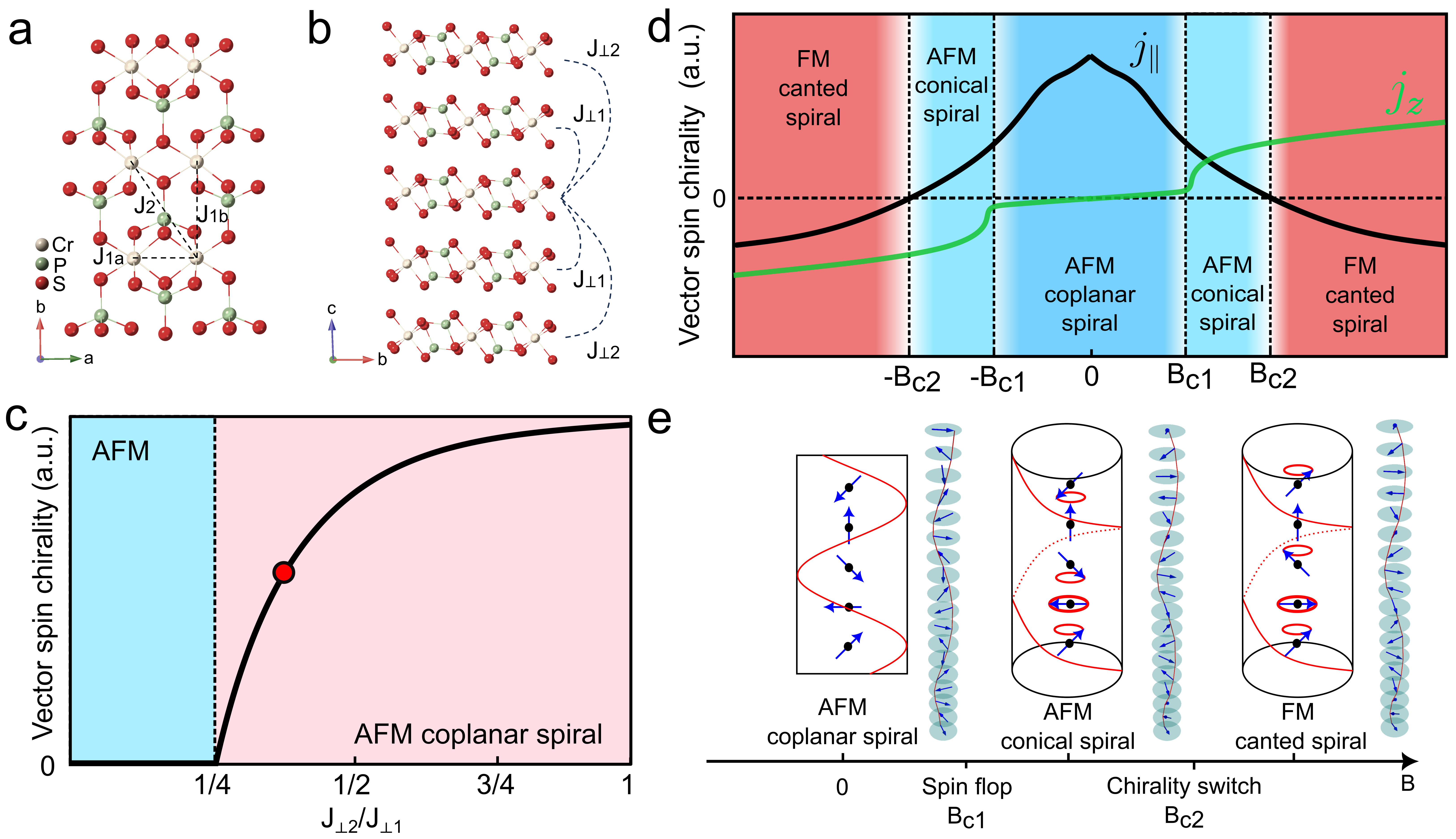}
    \caption{\textbf{Crystalline structure, exchange interaction model and predicted magnetic orders in CrPS$_4$.} (a) Schematic representation of the atomic structure of CrPS$_4$ with highlighted intralayer exchange interactions: $J_{1a}$, $J_{1b}$, and $J_{2}$. (b) Interlayer AFM exchange interactions (including nearest $J_{\perp1}$ and next-nearest $J_{\perp2}$ interlayer interactions), giving rise to the AFM coplanar spiral phase at zero field. (c) Schematic magnetic phase diagram as a function of the ratio of $J_{\perp2}/J_{\perp1}$. The red spot represents the theoretically calculated value of $J_{\perp2}/J_{\perp1}$ for CrPS$_4$. (d) Vector spin chirality as a function of the magnetic field, captured experimentally by the THG circular dichroism. Calculated even (in-plane, $j_\parallel$) and odd (out-of-plane, $j_z$) components are represented in black and light green, respectively. 
    The first critical field,~$\pm B_{c1}$, triggers the spin-flop transition from an AFM coplanar to an AFM conical spiral, while the second,~$\pm B_{c2}$, induces a chirality switch, marking the AFM conical to FM canted spiral transition. (e) Diagram of spin textures as a function of the magnetic field, illustrating spin flop and spin chirality switch: AFM coplanar spiral, AFM conical spiral, and FM canted spiral.}
    \label{fig:1theory}
\end{figure}

CrPS$_4$ is an air-stable vdW magnet known for its A-type antiferromagnetic (AFM) behavior, where ferromagnetic (FM) layers couple antiferromagnetically along the c-axis~\cite{Son2021,Wu2022,Seo2024,Budko2021}. The AFM order makes it an attractive choice for cryogenic spintronic devices, as unlike FM materials, antiferromagnets show zero net magnetization, leading to the absence of stray magnetic fields~\cite{Jungwirth2016}. However, as a consequence, external disturbances like small external magnetic fields, provide a relatively weak response in AFM materials, which poses challenges in probing spin behavior. Furthermore, despite growing interest, the theoretical understanding of CrPS$_4$ remains limited.
While there is a consensus regarding the AFM nature of the interlayer exchange interactions along the c-axis, the potential existence of competing couplings has remained an open question.

In terms of its crystalline structure (see Fig.~\ref{fig:1theory}a), monolayer CrPS$_4$ comprises CrS$_6$ edge-sharing octahedra, forming quasi-one-dimensional chains aligned along the b-axis. These chains are interconnected along the a-axis through PS$_4$ tetrahedra. The magnetic moments of CrPS$_4$ within each layer are placed in a rectangular lattice formed by the Cr$^{3+}$ ions. Spins within each layer are coupled to their neighbors by intralayer symmetric anisotropic FM exchange interactions (see Fig.~\ref{fig:1theory}a) and interlayer symmetric AFM exchange interactions (see Fig.~\ref{fig:1theory}b).

From first-principles density functional theory (DFT) methods (detailed in method and supplementary materials), the first ($J_{1a}$), second ($J_{1b}$) and third neighbor ($J_{2}$) exchange interactions within each layer
(see Fig.~\ref{fig:1theory}a) can be estimated. These intralayer couplings
lead to an FM order within each plane so that all the spins within a vdW plane remain collinear. In contrast, our DFT calculations show that a non-collinear spin spiral order between vdW planes emerges as a consequence of competition between nearest and next-nearest interlayer exchange interactions. Similar orders in other types of materials have been reported, for instance, as a consequence of spin-orbit coupling terms such as the interlayer Dzyaloshinskii-Moriya interaction \cite{Adachi1980,Sosnowska2002,Arkenbout2006}.
With this picture in mind, the spin-flop transition reported in bulk CrPS$_4$ stems from a metamagnetic transition associated with competing interlayer interactions, with the potential of featuring more complex phase transitions. To model this material, since the intralayer exchanges do not result in any additional frustration down to the monolayer limit, we can integrate out the degrees of freedom of individual spins within each layer, leading to a macro-spin Hamiltonian that reads


\begin{equation}
H_\perp = J_{\perp1}\sum_n \boldsymbol{S}_n\cdot\boldsymbol{S}_{n+1} + J_{\perp2}\sum_n\boldsymbol{S}_n\cdot\boldsymbol{S}_{n+2} + A_\parallel\sum_{n} \left[\left(S_{n}^x\right)^2+\left(S_{n,i}^y\right)^2\right]
\end{equation}

The nearest ($J_{\perp1}$) and next-nearest ($J_{\perp2}$) interlayer exchange interactions in CrPS$_4$ (as shown in Fig.~\ref{fig:1theory}b) are estimated with DFT calculations indicating a ratio $J_{\perp2}/J_{\perp1}$ well above the necessary threshold for non-collinear magnetism (see Fig.~\ref{fig:1theory}c). As known from the theory of frustrated magnets, competing AFM interactions between the first $J_{\perp1}$ and second neighbors $J_{\perp2}$ in a spin-chain can stabilize a spin-spiral phase in the event that $J_{\perp2}/J_{\perp1}>1/4$. Such a ground state can be determined by minimizing the energy of the spin Hamiltonian, from which we can confirm the presence of a spin-spiral phase.

Rationalizing this emergent magnetic phase
requires quantifying the non-collinearity of the magnetic texture, thereby characterizing both its chirality and periodicity. 
This can be achieved by making use of the average vector spin chirality, defined as $\boldsymbol{j}=\langle\boldsymbol{S}_n\times\boldsymbol{S}_{n+1}\rangle$~\cite{Menzel2012,KNB2005}. We shall see that this quantity evolves under a magnetic field in a manner compatible with the measured chiral THG response. The in-plane vector spin chirality $j_\parallel=\pm \sqrt{j_x^2+j_y^2}$ together with out-of-plane component $j_z$ then serve as the order parameters to capture both in-plane chirality switches as well as spin-flop transitions. It is worth noting that commensurability effects between the number of layers in the CrPS$_4$ samples, and the periodicity of the spiral induced by the $J_{2\perp}/J_{1\perp}$ ratio play a substantial role in determining the magnetic order and vector spin chirality of the material (see method and supplementary materials for details). The experimental observations from CrPS$_4$ are accounted for by taking~$J_{1\perp}\approx0.2$~meV, and~$J_{2\perp}\approx0.07$~meV, resulting in a ratio of $J_{2\perp}/J_{1\perp}\approx0.35$ (indicated by the red spot in Fig.~\ref{fig:1theory}c). 

We can therefore rationalize the effects of a magnetic field on few-layer CrPS$_4$ from the perspective of the vector spin chirality as shown in Fig.~\ref{fig:1theory}d with estimated order parameters (i.e., $j_\parallel$ and $j_z$). Without a magnetic field, this material displays a spontaneous finite vector spin chirality due to the competing interactions, reflecting a chiral magnetic texture driven by the competition between $J_{\perp1}$ and $J_{\perp2}$. This corresponds to a coplanar spin spiral phase. Beyond the spin-flop critical field ($B_{c1}$), the canting of spins drives a transition from the spontaneous AFM coplanar spiral phase to an AFM conical spiral phase. In this state, the spins adopt a helical arrangement, forming a screw-like structure that gives rise to an out-of-plane chiral spin texture. When the applied magnetic field exceeds $B_{c2}$, a further transition occurs from the AFM conical spiral phase to the FM canted spiral phase, accompanied by a chirality switch. Fig.~\ref{fig:1theory}e presents a schematic representation of the predicted diverse spin spiral phases, highlighting the spin-flop transition and subsequent switch in spin chirality. In the AFM conical spiral phase, adjacent spins are nearly anti-parallel, resulting in a characteristic sign for the spin cross-product $\textbf{S}_i \times \textbf{S}_{i+1}$. As the system transitions to the FM canted spiral phase, where spins become more aligned, this sign reverses. Consequently, the change in $j_{\parallel}$ reflects the realignment of spins, leading to a spin chirality switch. 

\section*{Chiral THG induced by non-collinear spins}

\begin{figure}[htbp]
    \centering
    \includegraphics[width=1\textwidth]{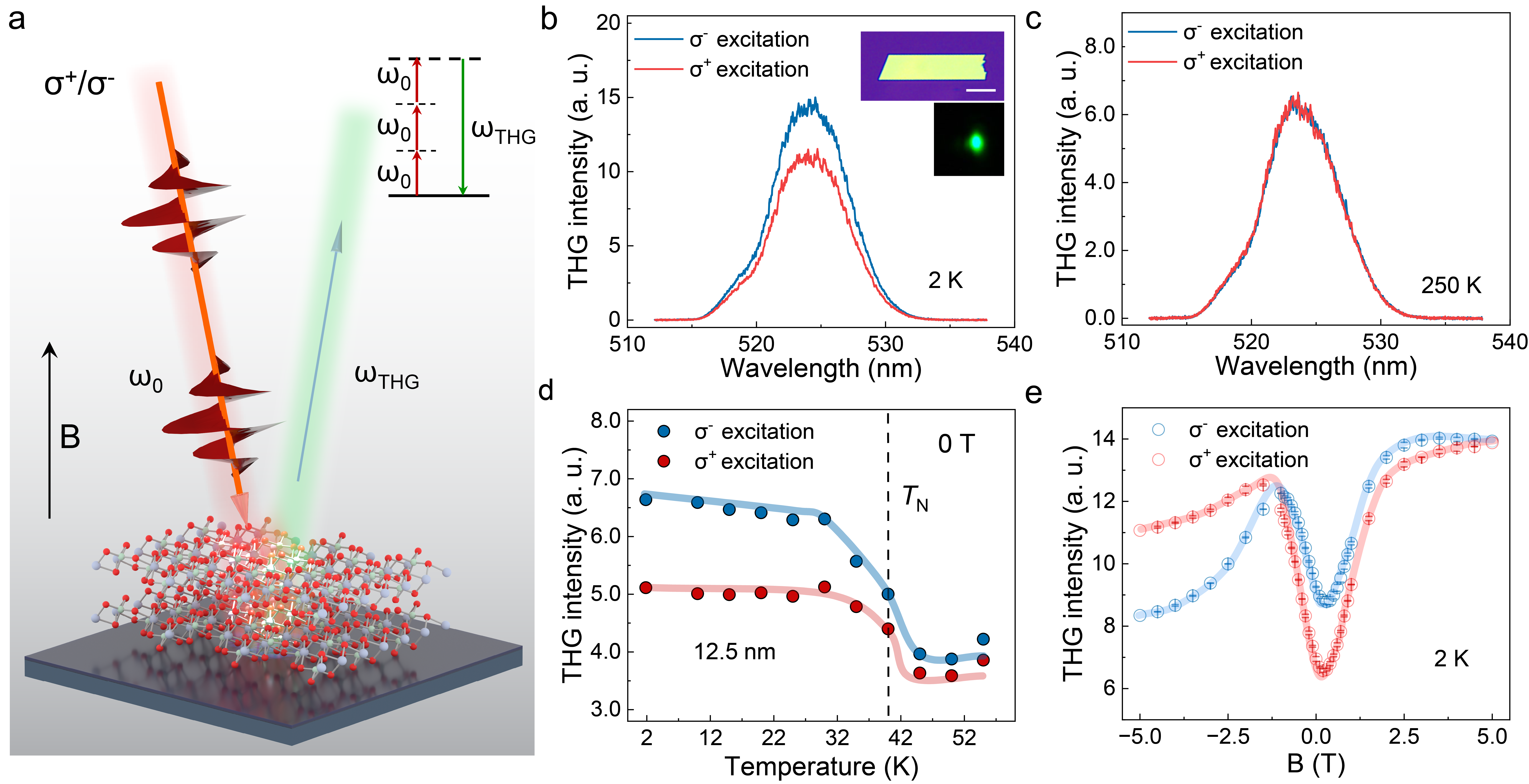}
    \caption{\textbf{Chiral THG in CrPS$_4$.} (a) Illustration of chiral THG measurement. $\omega_0$: the fundamental frequency, $\omega_\text{THG}$: the frequency of THG. Inset: schemtic THG process. THG spectra of a~$\sim$12~nm thick flake pumped with different chiralities at the temperature of (b)~2~K and (c)~250~K. Insets display the optical image of the CrPS$_4$ flake and the THG beam obtained by 1575~nm laser excitation, as captured by an imaging camera. Scale bar is $5~\mu\text{m}$. (d) Temperature-dependent THG intensity of the~$\sim$12~nm thick flake. $T_N$ indicates the Néel temperature~$\sim40$~K. (e) Out-of-plane magnetic field dependence of the chiral THG intensity at the temperature of~2~K.} 
    \label{fig:1exp}
\end{figure}

In our experiments, we investigate the non-collinear spin properties of CrPS$_4$ by measuring THG using a circularly polarized incident laser beam with different chiralities, as shown in Fig.~\ref{fig:1exp}a. Details of experimental methods and sample characterization are given in the method and supplementary materials. Fig.~\ref{fig:1exp}b displays the THG spectra from a~$\sim12$~nm thick CrPS$_4$ flake (thickness identified in supplementary Fig.~3), excited by a~$\sim1575$~nm femtosecond beam with different circular polarizations at the temperature of~$\sim2$~K. The THG intensity generated by the left-handed circularly polarized ($\sigma^-$) pump beam is higher ($\sim1.4$~times) than that produced by the right-handed circularly polarized ($\sigma^+$) beam. The insets show an optical image of the CrPS$_4$ flake and the corresponding strong THG signal at 1575~nm laser excitation. In contrast, at the higher temperature of~250~K, the two THG intensities equalize (Fig.~\ref{fig:1exp}c). Fig.~\ref{fig:1exp}d presents temperature-dependent THG measurements over~$\sim$2–55~K. The THG intensity decreases for both $\sigma^-$ and $\sigma^+$ light excitation as the temperature increases. Above~$\sim40$~K, the difference in THG intensity between the two chiralities becomes relatively small. The temperature of $\sim40$~K fits well with the previously published AFM transition at the Néel temperature ($T_N$~$\approx$~40~K) of CrPS$_4$~\cite{Peng2020}, indicating the correlation between chiral THG responses and the spin properties in CrPS$_4$.

To further investigate the correlation between chiral THG and spin properties in CrPS$_4$, we conduct out-of-plane magnetic field-dependent chiral THG measurements, as shown in Fig.~\ref{fig:1exp}e. When an upward magnetic field is applied ($B>0$~T), the THG intensities excited by both $\sigma^-$ and $\sigma^+$ beams increase and become nearly identical. However, under a downward magnetic field ($B<0$~T), both THG intensities initially increase as the magnetic field rises to $\sim-1.2$~T, after which they begin to decrease. The THG intensity excited by the $\sigma^+$ light exhibits a more rapid increase initially (red curve in Fig.~\ref{fig:1exp}e) until $\sim-1.2$~T, after which it decreases at a slower rate compared to the $\sigma^-$ light excitation. We confirm that the pronounced magnetic field-dependent chiral THG does not stem from the reflection circular dichroism (see supplementary Fig. 8 and 9), but instead arises from the THG response intrinsic to the spin texture of the CrPS$_4$ flake, emphasizing a robust chirality-dependent nonlinear optical interaction driven by the material's intrinsic chiral magnetic properties. In particular, the results suggest that an emergent spin spiral phase breaks spatial inversion symmetry by spontaneously generating a non-centrosymmetric magnetic structure. Further, the handedness of the spiral, captured by the vector spin chirality, introduces asymmetry in the interaction with circularly polarized light through a nonlinear process, allowing for a strong chiroptical THG response.

\section*{Probing spin phase transition with chiral THG}

\begin{figure}[htbp]
    \centering
    \includegraphics[width=0.8\textwidth]{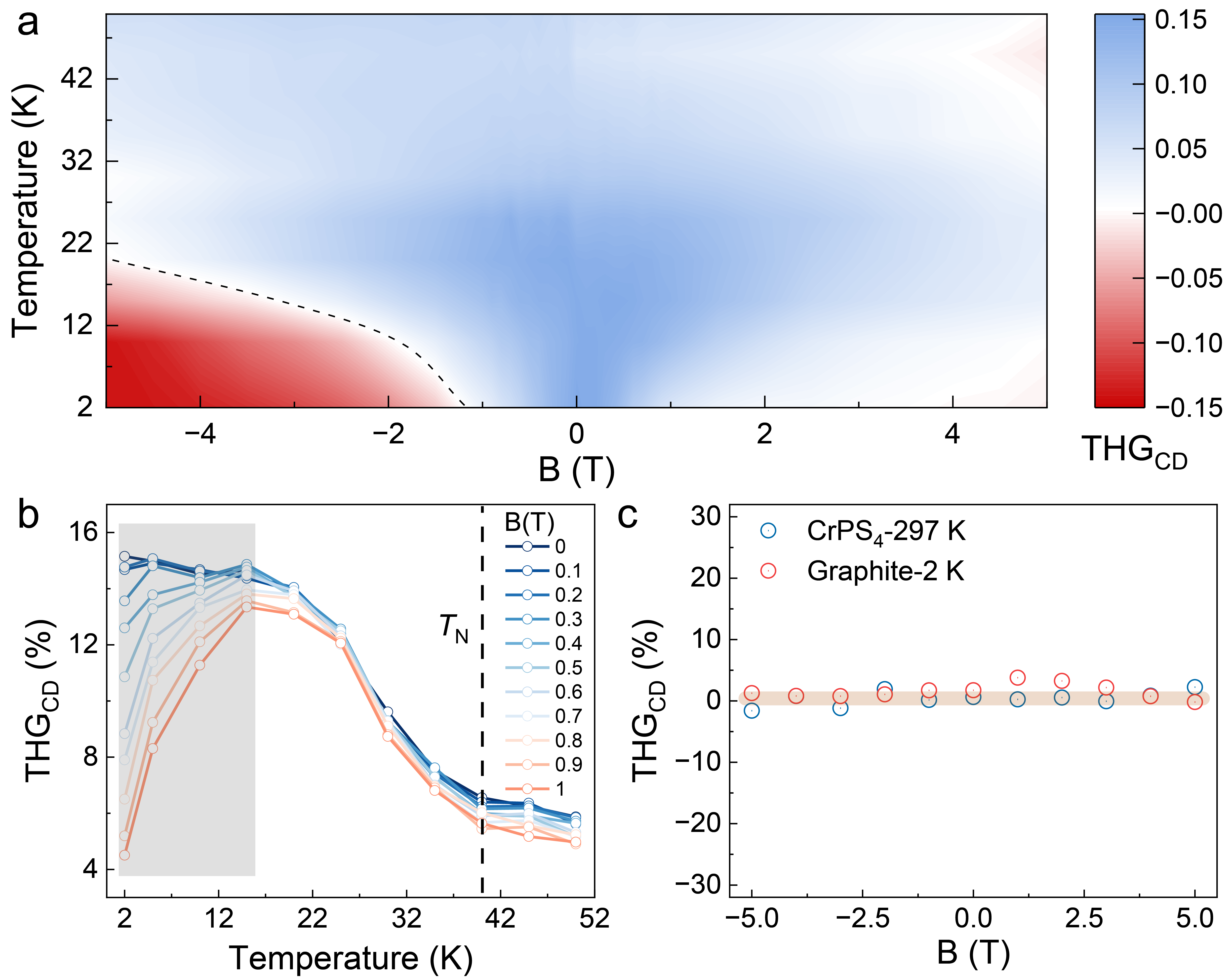}
    \caption{\textbf{Magnetic field and temperature-dependent THG circular dichroism in CrPS$_4$.} (a) Mapping results in the~$\sim12$~nm thick CrPS$_4$ flake. Dotted lines are plotted at points where THG$_\text{CD}$ equals zero to guide the eye, and they denote the optical chirality switch occurs. (b) Temperature-dependent THG$_\text{CD}$ under the $B$ field of~0~to~1~T. (c) Magnetic field dependent THG$_\text{CD}$ from the same CrPS$_4$ flake at 297 K and a graphite reference sample at~2~K.}    
    \label{fig:2exp}
\end{figure}

To gain deeper insights into the underlying magnetic properties using chiral THG, we utilize the THG circular dichroism (THG$_\text{CD}$), defined as \begin{math} \text{THG}_\text{CD}=\frac{I^- -I^+}{I^-+I^+} \end{math},
where $I^-$ represents the THG intensity excited by $\sigma^-$ light, and $I^+$ corresponds to the THG intensity generated by $\sigma^+$ light. Using data obtained at 2~K in Fig.~\ref{fig:1exp}b, the calculated THG$_\text{CD}$ is $\sim$14.5$\%$ for the $\sim$12~nm thick flake. Due to the negligible power dependence of THG$_\text{CD}$ shown in supplementary Fig. 10, we fix the power at 10~mW for all measurements to ensure consistency. We also perform THG\(_\text{CD}\) mapping, which confirms that THG$_\text{CD}$ primarily originates from the flake rather than the substrate (see supplementary Fig.~11). Similar THG$_\text{CD}$ signals are also observed in flakes with different thicknesses (from $\sim$12 to 80~nm in supplementary Fig.~12). The THG$_\text{CD}$ of the $\sim$12~nm thick flake is further measured as a function of temperature and magnetic field, as shown in Fig.~\ref{fig:2exp}a. Notably, the THG$_\text{CD}$ values range from $\sim-$0.14 to 0.15 for the flake. For clarity, we focus on measurements taken with a magnetic field from $-$5~to~5~T and temperatures from ~2~to~50~K, as these ranges exhibit the most appreciable changes in THG$_\text{CD}$ (supplementary Fig.~13). Under an upward magnetic field ($B>0$~T), THG$_\text{CD}$ decreases and approaches zero. Conversely, when a downward magnetic field is applied ($B<0$~T), the flake exhibits sign changes in THG$_\text{CD}$ (highlighted by dotted lines in Fig.~\ref{fig:2exp}a). For instance, at~2~K, THG$_\text{CD}$ transitions from positive to negative at $B\approx-1.2$~T, which aligns with the turning point observed in Fig.~\ref{fig:1exp}e at $B\approx-1.2$~T.

To emphasize the influence of the magnetic field, Fig.~\ref{fig:2exp}b provides detailed plots of the temperature and magnetic field dependence of THG$_\text{CD}$. The results reveal that the magnetic field significantly impacts the THG$_\text{CD}$ below $\sim$15~K (marked in the gray-shaded area). As the temperature approaches the Néel temperature ($T_N$, $\sim$40~K), the THG$_\text{CD}$ signal diminishes rapidly at different magnetic fields and becomes relatively constant for temperatures above $T_N$. We also measured THG$_\text{CD}$ from CrPS$_4$ at room temperature and from graphite at~2~K. In both cases, no significant changes were observed under different magnetic fields (Fig.~\ref{fig:2exp}c). These findings demonstrate that the chiral third-order nonlinear optical response of CrPS$_4$ is strongly influenced by temperature and external magnetic fields, a behavior uniquely associated with the chiral spin textures in the material.

\section*{Correlating tunable chiral spin textures with nonlinear optics}

\begin{figure}[htbp]
    \centering
    \includegraphics[width=1\textwidth]{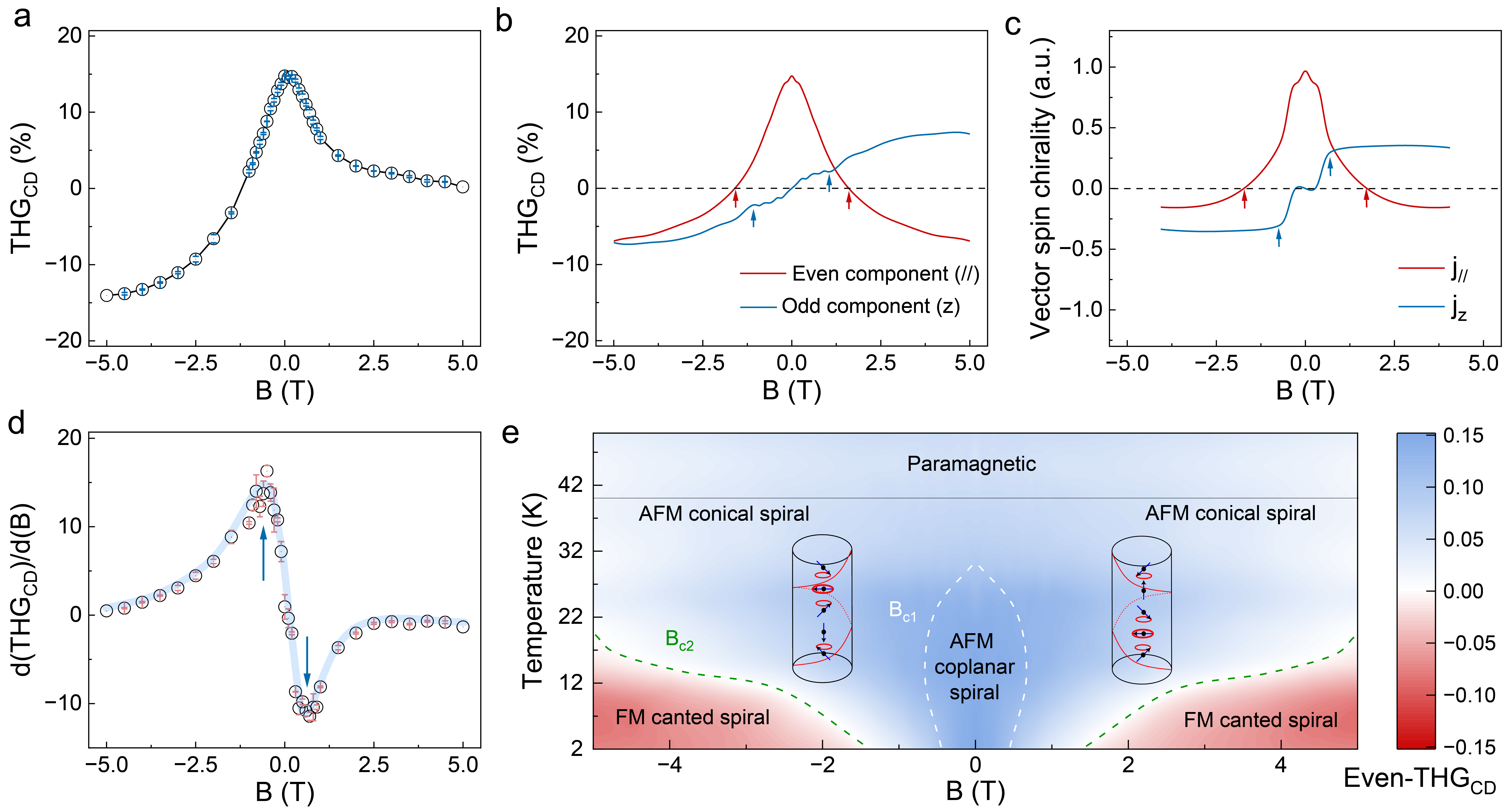}
    \caption{\textbf{Correlation between chiral THG measurement with the spin-feather structures in CrPS$_4$.} (a) Experimental signal of THG$_\text{CD}$ for the~$\sim12$~nm thick sample vs. $B$ field varying from~-5~to~5~T at~2~K. (b) Decomposition of the THG$_\text{CD}$ into even and odd components. Blue arrows mark changes in the behavior of the odd component, consistent with the predicted spin-flop transition at field $B_{c1}$, while red arrows mark sign switches in the even component of THG$_\text{CD}$ at critical field $B_{c2}$. (c) Predicted in-plane ($j_\parallel$) and out-of-plane ($j_z$) vector spin chirality averaged over an 18-layer (i.e., $\sim12$~nm) flake. A clear analogy is verified between the odd and even components of the THG$_\text{CD}$ and the vector spin chiralities, which serve here as a measure of the flake's chirality and non-collinearity. (d) Derivative of the THG$_\text{CD}$ data shown in (a). The blue arrows indicate a rapid change in THG$_\text{CD}$, which is consistent with the theory and closely aligns with the previously reported spin-flop transition in the bulk material \cite{Peng2020}. (e) Mapping of even components of THG$_\text{CD}$ as a function of $B$ field and temperature. The dashed white and green lines correspond to the critical field $\pm B_{c1}$ and $\pm B_{c2}$, respectively.}    
    \label{fig:2theory}
\end{figure}

We now draw the correlations between the THG measurement and the chiral spin structures in CrPS$_4$. We first plot the THG$_\text{CD}$ as a function of the magnetic field at~2~K in Fig.~\ref{fig:2theory}a. When an upward magnetic field is applied ($B>0$~T), the THG$_\text{CD}$ signal decreases and becomes close to 0. However, under a downward magnetic field ($B<0$~T), the THG$_\text{CD}$ starts to drop with the magnetic field and becomes zero at $\sim-1.2$~T, and continues to drop to $\sim-14\%$ at~-5~T. These results suggest a chirality switch under a magnetic field. Moreover, flipping the flake additionally offers the method of reversing its chiroptical response (see supplementary Fig.~14). From the perspective of the magnetic order, the in-plane vector spin chirality together with the out-of-plane component serve as the order parameters to capture both the spin chirality switch as well as the spin-flop transition. Both quantities can be experimentally accessed by examining the even and odd components of the chiroptical THG response of the material, as seen in Fig.~\ref{fig:2theory}. For better comparison, we extract the even and odd components of the chiroptical response of CrPS$_4$ in Fig.~\ref{fig:2theory}b. Fig.~\ref{fig:2theory}c shows the calculation of the in-plane, out-of-plane vector spin chirality under a magnetic field.

Fig.~\ref{fig:2theory}b and Fig.~\ref{fig:2theory}c showcase significant correlations between the extracted even and odd components of THG$_\text{CD}$ and the spin chirality. A direct comparison of the curves plotted in Fig.~\ref{fig:2theory} b and c suggests that the chiroptical THG response can be well captured by the behavior of the vector spin chirality. The asymmetrical curve regarding the experimentally applied magnetic field corresponds to the mixing of contributions stemming from in-plane vector spin chirality $j_\parallel$ and out-of-plane component $j_z$. The behavior of $\boldsymbol{j}$ can be understood as follows: As the external magnetic field is increased from $B=0$~T, the oscillations in spin along the different layers in the $c$-axis direction tend to align preferentially with the field. Without an external field, the spins oscillate along the layers, reaching the maximum and minimum allowed values for the projection in the out-of-plane direction $S^z=\pm3/2$ and all lie in a single plane. The magnetic field results in an effective decrease in non-collinearity, which is quantified by an overall decrease in the in-plane vector spin chirality $j_\parallel$ caused by a rearrangement of the oscillations. This causes an overall preference towards either the positive or negative direction along the c-axis, depending on the direction of the applied magnetic field. On the other hand, $j_z$ remains close to zero in this first regime, while the decrease in $j_\parallel$ is accompanied by a subtle magnetization increase. At a value of magnetic field $B_{c1}\approx 0.49$~T, a sharp increase in the $j_z$ reflects the occurrence of the spin-flop transition. This is accompanied by a sharp increase in $j_z$ to a finite value, which can be positive or negative depending on the field direction. This behavior, since the spin configuration no longer attains a maximum and minimum projection $S^z=\pm3/2$, is no longer coplanar. Instead, it stabilizes a sort of AFM conical spiral phase, with $S^z$ oscillating symmetrically around a finite positive value. Continuously increasing the magnetic field magnitude above the spin-flop cants all the spins towards the same direction with the average sign of the in-plane vector spin chirality $j_\parallel$ being switched at a critical magnetic field $B_{c2}\approx1.5$~T. For much larger fields than those considered here, the canting culminates in an FM phase, with the Zeeman coupling dominating over all other sources of magnetic frustration. Thus, two magnetic phase transitions are observed, the first corresponding to a fast decrease in $j_\parallel$ and a sharp increase in $j_z$, and the second to a sign switch in $j_\parallel$. To gain more insight into these transitions, we can also compute the derivative of the THG$_\text{CD}$ data relative to the magnetic field (Fig.~\ref{fig:2theory}d) at~2~K. Two distinct transition points (marked by blue arrows), indicating fast changes in the magnetic order, are observed at approximately $\pm0.5$~T, which align closely with the spin-flop transition in CrPS$_4$ bulk materials at low temperatures~\cite{Peng2020}. Fig.~\ref{fig:2theory}e illustrates the various magnetic phases at finite temperature with our experimental data, including the paramagnetic state, AFM coplanar spiral, AFM conical spiral, and FM canted spiral. Notably, the FM canted spiral phase is realized when the even component of THG$_\text{CD}$ exhibits a sign change. This phase remains stable beyond the critical field $B_{c2}$; however, as the temperature increases, thermal fluctuations destabilize the spin configuration, necessitating a stronger magnetic field to sustain the FM canted spiral state. These results demonstrate that the THG$_\text{CD}$ is sensitive to magnetic phase transitions, providing insights into the underlying tunable spin behavior.

\section*{Conclusions}

In conclusion, we have observed spontaneous chiral AFM spin textures in the vdW magnet CrPS$_4$ through the correlation between its optical and magnetic chiral interaction. Our findings demonstrate that CrPS$_4$ supports a diverse range of chiral magnetic structures, including AFM coplanar spiral, AFM conical and FM canted spin spirals. These chiral behaviors are highly tunable, governed by the interplay between the interlayer and intralayer spin coupling under applied magnetic fields.
Moreover, our chiral THG experiments unveil the evolution of chirality driven by complex spin spiral structures under varying magnetic fields and temperatures. This study demonstrates a highly sensitive and efficient method for probing spin chirality, shedding light on intricate spin coupling interactions. Beyond its significance as a powerful experimental tool for fundamental research, our approach also accelerates technological advancements in spin-photonics and spintronics.
Our findings establish CrPS$_4$ and its heterostructures as a unique platform for exploring spin chirality in emerging 2D magnets, where leveraging advanced 2D device fabrication techniques and stacking engineering, the interplay between intra- and interlayer exchange interactions, magnetic order, and light-matter coupling could be effectively manipulated to enable innovative applications.
Ultimately, the optical probing of spin chirality provides a new strategy to understand chiral magnetic phenomena, paving the way to unlock opportunities for the development of high-speed, energy-efficient chiral devices. 

\section*{Methods}

\subsection*{Sample preparation and characterization}

Flux zone-grown CrPS$_4$ crystals are purchased from 2D Semiconductor. CrPS$_4$ flakes are mechanically exfoliated by the dry transfer technique using a PDMS stamp to silicon chips with 285~nm thick SiO$_2$. The thickness of the samples is characterized using atomic force microscopy with a Dimension Icon system from Bruker. The Raman spectrum is measured by a WITEC alpha 300 RA+ system at room temperature. A 532~nm continuous wave laser is focused by a 100X CF Plan Nikon objective (NA~=~0.95), and the reflected Raman signal is collected by the same objective and sent to a spectrometer. 

\subsection*{Chiral THG measurement}

Circular polarization-dependent optical THG measurements at different temperatures are carried out using a custom-built confocal optical microscope, operating in reflection geometry (see supplementary Fig.~2 for details). The sample is mounted into a closed-loop cryostat (attoDRY 2100) equipped with superconducting magnets (-9 to 9~T). The femtosecond laser at 1575~nm (FemtoFiber smart 780) is focused onto the sample by a low-temperature objective from Attocube (LT-APO/Telecom, NA~=~0.8). A quarter-wave plate mounted on a motorized rotation stage (KPRM1E/M) is employed to rotate the chirality of the pump light. The reflected THG signal is collected by the same objective. A dichroic mirror with neglect polarization dependence is utilized to separate the signal. Then, the signal is further coupled into a multimode fiber to an Andor Shamrock 750 spectrograph equipped with an electron multiply charge-coupled device (EMCCD, Newton 970).

\subsection*{Theoretical model and calculations}

CrPS$_4$ realizes a spin-$3/2$ Heisenberg model, with an easy axis
associated with a positive in-plane single-ion anisotropy.
The Hamiltonian for such a system in the presence of an external out-of-plane magnetic field takes the form


\begin{equation}
H = -\sum_{i,j,n}J_{i,j} \boldsymbol{S}_{n,i}\cdot\boldsymbol{S}_{n,j} + \sum_{n,m,i}J_{\perp n,m}\boldsymbol{S}_{n,i}\cdot\boldsymbol{S}_{m,i} + A_\parallel\sum_{n,i}\left[\left(S_{n,i}^x\right)^2+\left(S_{n,i}^y\right)^2\right] + g\mu_BB\sum_{i,n} S_{n,i}^z
\label{eq:usual_Heisenberg}
\end{equation}

Here, the indices $i$ and $j$ run over each in-plane spins whereas $n$ and $m$ run over different layers. As such, $J_{i,j}$ represents the in-plane exchange interaction between sites $i$ and $j$.
On the other hand, $J_{\perp n,m}$ accounts for out-of-plane interlayer exchange interactions promoting an out-of-plane antiferromagnetic order. 

To quantitatively account for the experimentally observed chiroptical measurements, we extend the previous works to also account for the next nearest interlayer interactions as well as include the effects of intralayer anisotropic exchanges.

In particular, we include the effect of intralayer exchange interactions up to third nearest neighbors in few-layer CrPS$_4$ ($J_{1a}$, $J_{1b}$, and $J_{2}$ in Fig.~
\ref{fig:1theory}a), and compute the magnetic ground state for a rectangular lattice of Cr$^{3+}$ ions.


\begin{equation}
H_\parallel^{(n)} = J_{1a}\sum_{\langle i,j\rangle}\boldsymbol{S}_{n,i}\cdot\boldsymbol{S}_{n,j} + J_{1b}\sum_{\langle\langle i,j\rangle\rangle}\boldsymbol{S}_{n,i}\cdot\boldsymbol{S}_{n,j} + J_2\sum_{\langle\langle\langle i,j\rangle\rangle\rangle}\boldsymbol{S}_{n,i}\cdot\boldsymbol{S}_{n,j} + A_\parallel\sum_{i}\left[\left(S_{n,i}^x\right)^2+\left(S_{n,i}^y\right)^2\right]
\end{equation}

where $H^{(n)}_\parallel$ stands for the intralayer Hamiltonian of the $n^\text{th}$ layer, such that the total Hamiltonian is $H=\sum_n H_\parallel^{(n)} + H_\perp$. The term $H_\perp$ corresponds to the interlayer component of the Hamiltonian for few-layer CrPS$_4$. Taking into account exchange couplings between the nearest and next-nearest layers, it reads

\begin{equation}
H_\perp=J_{\perp1}\sum_{i,n} \boldsymbol{S}_{n,i}\cdot\boldsymbol{S}_{n+1,i} + J_{\perp2}\sum_{i,n} \boldsymbol{S}_{n,i}\cdot\boldsymbol{S}_{n+2,i}
\end{equation}

With this Heisenberg model, the different magnetic transitions in few layer CrPS$_4$ are directly captured. In particular, in order to estimate the values, including the effects of the intralayer and interlayer interactions on the spin ground state, we have performed \textit{ab initio} Density Functional Theory (DFT) calculations with the all-electron full-potential linearized augmented-plane-wave method as implemented in the Elk code \cite{elk}.

\bibliography{sample}

\begin{thebibliography}{10}
\urlstyle{rm}
\expandafter\ifx\csname url\endcsname\relax
  \def\url#1{\texttt{#1}}\fi
\expandafter\ifx\csname urlprefix\endcsname\relax\def\urlprefix{URL }\fi
\expandafter\ifx\csname doiprefix\endcsname\relax\def\doiprefix{DOI: }\fi
\providecommand{\bibinfo}[2]{#2}
\providecommand{\eprint}[2][]{\url{#2}}

\bibitem{Gibertini2019}
\bibinfo{author}{Gibertini, M.}, \bibinfo{author}{Koperski, M.}, \bibinfo{author}{Morpurgo, A.~F.} \& \bibinfo{author}{Novoselov, K.~S.}
\newblock \bibinfo{journal}{\bibinfo{title}{Magnetic 2\textrm{D} materials and heterostructures}}.
\newblock {\emph{\JournalTitle{Nat. Nanotechnol.}}} \textbf{\bibinfo{volume}{14}}, \bibinfo{pages}{408–419} (\bibinfo{year}{2019}).
\newblock \doiprefix\url{10.1038/s41565-019-0438-6}.

\bibitem{Zhang2024}
\bibinfo{author}{Zhang, B.}, \bibinfo{author}{Lu, P.}, \bibinfo{author}{Tabrizian, R.}, \bibinfo{author}{Feng, P. X.-L.} \& \bibinfo{author}{Wu, Y.}
\newblock \bibinfo{journal}{\bibinfo{title}{2\textrm{D} magnetic heterostructures: spintronics and quantum future}}.
\newblock {\emph{\JournalTitle{npj Spintron.}}} \textbf{\bibinfo{volume}{2}}, \bibinfo{pages}{6} (\bibinfo{year}{2024}).
\newblock \doiprefix\url{10.1038/s44306-024-00011-w}.

\bibitem{Blei2021}
\bibinfo{author}{Blei, M.} \emph{et~al.}
\newblock \bibinfo{journal}{\bibinfo{title}{Synthesis, engineering, and theory of 2\textrm{D} van der \textrm{Waals} magnets}}.
\newblock {\emph{\JournalTitle{Appl. Phys. Rev.}}} \textbf{\bibinfo{volume}{8}}, \bibinfo{pages}{021301} (\bibinfo{year}{2021}).
\newblock \doiprefix\url{10.1063/5.0025658}.

\bibitem{Zhang2022_optical}
\bibinfo{author}{Zhang, P.} \emph{et~al.}
\newblock \bibinfo{journal}{\bibinfo{title}{All-optical switching of magnetization in atomically thin \textrm{CrI}$_3$}}.
\newblock {\emph{\JournalTitle{Nat. Mater.}}} \textbf{\bibinfo{volume}{21}}, \bibinfo{pages}{1373–1378} (\bibinfo{year}{2022}).
\newblock \doiprefix\url{10.1038/s41563-022-01354-7}.

\bibitem{Lee2016}
\bibinfo{author}{Lee, J.-U.} \emph{et~al.}
\newblock \bibinfo{journal}{\bibinfo{title}{Ising-type magnetic ordering in atomically thin \textrm{FePS}$_3$}}.
\newblock {\emph{\JournalTitle{Nano Lett.}}} \textbf{\bibinfo{volume}{16}}, \bibinfo{pages}{7433–7438} (\bibinfo{year}{2016}).
\newblock \doiprefix\url{10.1021/acs.nanolett.6b03052}.

\bibitem{Gong_2017}
\bibinfo{author}{Gong, C.} \emph{et~al.}
\newblock \bibinfo{journal}{\bibinfo{title}{Discovery of intrinsic ferromagnetism in two-dimensional van der \textrm{Waals} crystals}}.
\newblock {\emph{\JournalTitle{Nature}}} \textbf{\bibinfo{volume}{546}}, \bibinfo{pages}{265–269} (\bibinfo{year}{2017}).
\newblock \doiprefix\url{10.1038/nature22060}.

\bibitem{Huang_2017}
\bibinfo{author}{Huang, B.} \emph{et~al.}
\newblock \bibinfo{journal}{\bibinfo{title}{Layer-dependent ferromagnetism in a van der \textrm{Waals} crystal down to the monolayer limit}}.
\newblock {\emph{\JournalTitle{Nature}}} \textbf{\bibinfo{volume}{546}}, \bibinfo{pages}{270–273} (\bibinfo{year}{2017}).
\newblock \doiprefix\url{10.1038/nature22391}.

\bibitem{Cai2019}
\bibinfo{author}{Cai, X.} \emph{et~al.}
\newblock \bibinfo{journal}{\bibinfo{title}{Atomically thin \textrm{CrCl}$_3$: an in-plane layered antiferromagnetic insulator}}.
\newblock {\emph{\JournalTitle{Nano Lett.}}} \textbf{\bibinfo{volume}{19}}, \bibinfo{pages}{3993–3998} (\bibinfo{year}{2019}).
\newblock \doiprefix\url{10.1021/acs.nanolett.9b01317}.

\bibitem{OFumega2020_Crmagnets}
\bibinfo{author}{O.~Fumega, A.} \emph{et~al.}
\newblock \bibinfo{journal}{\bibinfo{title}{Electronic structure and magnetic exchange interactions of \textrm{Cr}-based van der \textrm{Waals} ferromagnets. \textrm{A} comparative study between \textrm{CrBr}$_3$ and \textrm{Cr$_2$Ge$_2$Te$_6$}}}.
\newblock {\emph{\JournalTitle{J. Mater. Chem. C}}} \textbf{\bibinfo{volume}{8}}, \bibinfo{pages}{13582–13589} (\bibinfo{year}{2020}).
\newblock \doiprefix\url{10.1039/d0tc02003f}.

\bibitem{Sun2021}
\bibinfo{author}{Sun, Q.-C.} \emph{et~al.}
\newblock \bibinfo{journal}{\bibinfo{title}{Magnetic domains and domain wall pinning in atomically thin \textrm{CrBr}$_3$ revealed by nanoscale imaging}}.
\newblock {\emph{\JournalTitle{Nat. Commun.}}} \textbf{\bibinfo{volume}{12}}, \bibinfo{pages}{1989} (\bibinfo{year}{2021}).
\newblock \doiprefix\url{10.1038/s41467-021-22239-4}.

\bibitem{Yao2023}
\bibinfo{author}{Yao, F.} \emph{et~al.}
\newblock \bibinfo{journal}{\bibinfo{title}{Multiple antiferromagnetic phases and magnetic anisotropy in exfoliated \textrm{CrBr}$_3$ multilayers}}.
\newblock {\emph{\JournalTitle{Nat. Commun.}}} \textbf{\bibinfo{volume}{14}}, \bibinfo{pages}{4969} (\bibinfo{year}{2023}).
\newblock \doiprefix\url{10.1038/s41467-023-40723-x}.

\bibitem{Fert_2013}
\bibinfo{author}{Fert, A.}, \bibinfo{author}{Cros, V.} \& \bibinfo{author}{Sampaio, J.}
\newblock \bibinfo{journal}{\bibinfo{title}{Skyrmions on the track}}.
\newblock {\emph{\JournalTitle{Nat. Nanotechnol.}}} \textbf{\bibinfo{volume}{8}}, \bibinfo{pages}{152–156} (\bibinfo{year}{2013}).
\newblock \doiprefix\url{10.1038/nnano.2013.29}.

\bibitem{Zhang_2023}
\bibinfo{author}{Zhang, H.} \emph{et~al.}
\newblock \bibinfo{journal}{\bibinfo{title}{Magnetic skyrmions: materials, manipulation, detection, and applications in spintronic devices}}.
\newblock {\emph{\JournalTitle{Mater. Futures}}} \textbf{\bibinfo{volume}{2}}, \bibinfo{pages}{032201} (\bibinfo{year}{2023}).
\newblock \doiprefix\url{10.1088/2752-5724/ace1df}.

\bibitem{Back_2020}
\bibinfo{author}{Back, C.} \emph{et~al.}
\newblock \bibinfo{journal}{\bibinfo{title}{The 2020 skyrmionics roadmap}}.
\newblock {\emph{\JournalTitle{J. Phys. D: Appl. Phys.}}} \textbf{\bibinfo{volume}{53}}, \bibinfo{pages}{363001} (\bibinfo{year}{2020}).
\newblock \doiprefix\url{10.1088/1361-6463/ab8418}.

\bibitem{Wang_2022_skyrmions}
\bibinfo{author}{Wang, K.}, \bibinfo{author}{Bheemarasetty, V.}, \bibinfo{author}{Duan, J.}, \bibinfo{author}{Zhou, S.} \& \bibinfo{author}{Xiao, G.}
\newblock \bibinfo{journal}{\bibinfo{title}{Fundamental physics and applications of skyrmions: \textrm{A} review}}.
\newblock {\emph{\JournalTitle{J. Magn. Magn. Mater.}}} \textbf{\bibinfo{volume}{563}}, \bibinfo{pages}{169905} (\bibinfo{year}{2022}).
\newblock \doiprefix\url{10.1016/j.jmmm.2022.169905}.

\bibitem{Song2021}
\bibinfo{author}{Song, T.} \emph{et~al.}
\newblock \bibinfo{journal}{\bibinfo{title}{Direct visualization of magnetic domains and moiré magnetism in twisted 2\textrm{D} magnets}}.
\newblock {\emph{\JournalTitle{Science}}} \textbf{\bibinfo{volume}{374}}, \bibinfo{pages}{1140–1144} (\bibinfo{year}{2021}).
\newblock \doiprefix\url{10.1126/science.abj7478}.

\bibitem{Yang_2024}
\bibinfo{author}{Yang, B.} \emph{et~al.}
\newblock \bibinfo{journal}{\bibinfo{title}{Macroscopic tunneling probe of moiré spin textures in twisted \textrm{CrI}$_3$}}.
\newblock {\emph{\JournalTitle{Nat. Commun.}}} \textbf{\bibinfo{volume}{15}}, \bibinfo{pages}{4982} (\bibinfo{year}{2024}).
\newblock \doiprefix\url{10.1038/s41467-024-49261-6}.

\bibitem{Xie2023}
\bibinfo{author}{Xie, H.} \emph{et~al.}
\newblock \bibinfo{journal}{\bibinfo{title}{Evidence of non-collinear spin texture in magnetic moiré superlattices}}.
\newblock {\emph{\JournalTitle{Nat. Phys.}}} \textbf{\bibinfo{volume}{19}}, \bibinfo{pages}{1150–1155} (\bibinfo{year}{2023}).
\newblock \doiprefix\url{10.1038/s41567-023-02061-z}.

\bibitem{Song_2022}
\bibinfo{author}{Song, Q.} \emph{et~al.}
\newblock \bibinfo{journal}{\bibinfo{title}{Evidence for a single-layer van der \textrm{Waals} multiferroic}}.
\newblock {\emph{\JournalTitle{Nature}}} \textbf{\bibinfo{volume}{602}}, \bibinfo{pages}{601–605} (\bibinfo{year}{2022}).
\newblock \doiprefix\url{10.1038/s41586-021-04337-x}.

\bibitem{Amini_2024}
\bibinfo{author}{Amini, M.} \emph{et~al.}
\newblock \bibinfo{journal}{\bibinfo{title}{Atomic‐scale visualization of multiferroicity in monolayer \textrm{NiI}$_2$}}.
\newblock {\emph{\JournalTitle{Adv. Mater.}}} \textbf{\bibinfo{volume}{36}}, \bibinfo{pages}{202311342} (\bibinfo{year}{2024}).
\newblock \doiprefix\url{10.1002/adma.202311342}.

\bibitem{Fumega_2022}
\bibinfo{author}{Fumega, A.~O.} \& \bibinfo{author}{Lado, J.~L.}
\newblock \bibinfo{journal}{\bibinfo{title}{Microscopic origin of multiferroic order in monolayer \textrm{NiI}$_2$}}.
\newblock {\emph{\JournalTitle{2D Mater.}}} \textbf{\bibinfo{volume}{9}}, \bibinfo{pages}{025010} (\bibinfo{year}{2022}).
\newblock \doiprefix\url{10.1088/2053-1583/ac4e9d}.

\bibitem{Anto2024}
\bibinfo{author}{Antão, T. V.~C.}, \bibinfo{author}{Lado, J.~L.} \& \bibinfo{author}{Fumega, A.~O.}
\newblock \bibinfo{journal}{\bibinfo{title}{Electric field control \textrm{of} moiré skyrmion phases in twisted multiferroic \textrm{NiI}$_2$ bilayers}}.
\newblock {\emph{\JournalTitle{Nano Lett.}}} \textbf{\bibinfo{volume}{24}}, \bibinfo{pages}{15767–15773} (\bibinfo{year}{2024}).
\newblock \doiprefix\url{10.1021/acs.nanolett.4c04582}.

\bibitem{Li_2024}
\bibinfo{author}{Li, J.} \emph{et~al.}
\newblock \bibinfo{journal}{\bibinfo{title}{Signatures of polarized chiral spin disproportionation in rare earth nickelates}}.
\newblock {\emph{\JournalTitle{Nat. Commun.}}} \textbf{\bibinfo{volume}{15}}, \bibinfo{pages}{7427} (\bibinfo{year}{2024}).
\newblock \doiprefix\url{10.1038/s41467-024-51576-3}.

\bibitem{Lan2020}
\bibinfo{author}{Lan, T.}, \bibinfo{author}{Ding, B.} \& \bibinfo{author}{Liu, B.}
\newblock \bibinfo{journal}{\bibinfo{title}{Magneto‐optic effect of two‐dimensional materials and related applications}}.
\newblock {\emph{\JournalTitle{Nano Sel.}}} \textbf{\bibinfo{volume}{1}}, \bibinfo{pages}{298–310} (\bibinfo{year}{2020}).
\newblock \doiprefix\url{10.1002/nano.202000032}.

\bibitem{MolinaSnchez2020}
\bibinfo{author}{Molina-Sánchez, A.}, \bibinfo{author}{Catarina, G.}, \bibinfo{author}{Sangalli, D.} \& \bibinfo{author}{Fernández-Rossier, J.}
\newblock \bibinfo{journal}{\bibinfo{title}{Magneto-optical response of chromium trihalide monolayers: chemical trends}}.
\newblock {\emph{\JournalTitle{J. Mater. Chem. C}}} \textbf{\bibinfo{volume}{8}}, \bibinfo{pages}{8856–8863} (\bibinfo{year}{2020}).
\newblock \doiprefix\url{10.1039/d0tc01322f}.

\bibitem{Hendriks2021}
\bibinfo{author}{Hendriks, F.} \& \bibinfo{author}{Guimarães, M. H.~D.}
\newblock \bibinfo{journal}{\bibinfo{title}{Enhancing magneto-optic effects in two-dimensional magnets by thin-film interference}}.
\newblock {\emph{\JournalTitle{AIP Adv.}}} \textbf{\bibinfo{volume}{11}}, \bibinfo{pages}{035132} (\bibinfo{year}{2021}).
\newblock \doiprefix\url{10.1063/5.0040262}.

\bibitem{Kato2023}
\bibinfo{author}{Kato, Y.~D.}, \bibinfo{author}{Okamura, Y.}, \bibinfo{author}{Hirschberger, M.}, \bibinfo{author}{Tokura, Y.} \& \bibinfo{author}{Takahashi, Y.}
\newblock \bibinfo{journal}{\bibinfo{title}{Topological magneto-optical effect from skyrmion lattice}}.
\newblock {\emph{\JournalTitle{Nat. Commun.}}} \textbf{\bibinfo{volume}{14}}, \bibinfo{pages}{5416} (\bibinfo{year}{2023}).
\newblock \doiprefix\url{10.1038/s41467-023-41203-y}.

\bibitem{Suzuki2022}
\bibinfo{author}{Suzuki, M.} \emph{et~al.}
\newblock \bibinfo{journal}{\bibinfo{title}{Magnetic anisotropy of the van der \textrm{Waals} ferromagnet \textrm{Cr$_2$Ge$_2$Te$_6$} studied by angular-dependent x-ray magnetic circular dichroism}}.
\newblock {\emph{\JournalTitle{Phys. Rev. Res.}}} \textbf{\bibinfo{volume}{4}}, \bibinfo{pages}{013139} (\bibinfo{year}{2022}).
\newblock \doiprefix\url{10.1103/PhysRevResearch.4.013139}.

\bibitem{Ando1992}
\bibinfo{author}{Ando, K.}, \bibinfo{author}{Takahashi, K.}, \bibinfo{author}{Okuda, T.} \& \bibinfo{author}{Umehara, M.}
\newblock \bibinfo{journal}{\bibinfo{title}{Magnetic circular dichroism of zinc-blende-phase \textrm{MnTe}}}.
\newblock {\emph{\JournalTitle{Phys. Rev. B}}} \textbf{\bibinfo{volume}{46}}, \bibinfo{pages}{12289--12297} (\bibinfo{year}{1992}).
\newblock \doiprefix\url{10.1103/PhysRevB.46.12289}.

\bibitem{Burch2018}
\bibinfo{author}{Burch, K.~S.}, \bibinfo{author}{Mandrus, D.} \& \bibinfo{author}{Park, J.-G.}
\newblock \bibinfo{journal}{\bibinfo{title}{Magnetism in two-dimensional van der \textrm{Waals} materials}}.
\newblock {\emph{\JournalTitle{Nature}}} \textbf{\bibinfo{volume}{563}}, \bibinfo{pages}{47–52} (\bibinfo{year}{2018}).
\newblock \doiprefix\url{10.1038/s41586-018-0631-z}.

\bibitem{Harada2018}
\bibinfo{author}{Harada, Y.}, \bibinfo{author}{Haraguchi, E.}, \bibinfo{author}{Kaneshima, K.} \& \bibinfo{author}{Sekikawa, T.}
\newblock \bibinfo{journal}{\bibinfo{title}{Circular dichroism in high-order harmonic generation from chiral molecules}}.
\newblock {\emph{\JournalTitle{Phys. Rev. A}}} \textbf{\bibinfo{volume}{98}}, \bibinfo{pages}{021401} (\bibinfo{year}{2018}).
\newblock \doiprefix\url{10.1103/PhysRevA.98.021401}.

\bibitem{Gandolfi2021}
\bibinfo{author}{Gandolfi, M.}, \bibinfo{author}{Tognazzi, A.}, \bibinfo{author}{Rocco, D.}, \bibinfo{author}{De~Angelis, C.} \& \bibinfo{author}{Carletti, L.}
\newblock \bibinfo{journal}{\bibinfo{title}{Near-unity third-harmonic circular dichroism driven by a quasibound state in the continuum in asymmetric silicon metasurfaces}}.
\newblock {\emph{\JournalTitle{Phys. Rev. A}}} \textbf{\bibinfo{volume}{104}}, \bibinfo{pages}{023524} (\bibinfo{year}{2021}).
\newblock \doiprefix\url{10.1103/physreva.104.023524}.

\bibitem{Toftul2024}
\bibinfo{author}{Toftul, I.} \emph{et~al.}
\newblock \bibinfo{journal}{\bibinfo{title}{Chiral dichroism in resonant metasurfaces with monoclinic lattices}}.
\newblock {\emph{\JournalTitle{Phys. Rev. Lett.}}} \textbf{\bibinfo{volume}{133}}, \bibinfo{pages}{216901} (\bibinfo{year}{2024}).
\newblock \doiprefix\url{10.1103/PhysRevLett.133.216901}.

\bibitem{Kim_2020}
\bibinfo{author}{Kim, D.} \emph{et~al.}
\newblock \bibinfo{journal}{\bibinfo{title}{Giant nonlinear circular dichroism from intersubband polaritonic metasurfaces}}.
\newblock {\emph{\JournalTitle{Nano Lett.}}} \textbf{\bibinfo{volume}{20}}, \bibinfo{pages}{8032–8039} (\bibinfo{year}{2020}).
\newblock \doiprefix\url{10.1021/acs.nanolett.0c02978}.

\bibitem{Tonkaev2024}
\bibinfo{author}{Tonkaev, P.} \emph{et~al.}
\newblock \bibinfo{journal}{\bibinfo{title}{Nonlinear chiral metasurfaces based on structured van der \textrm{Waals} materials}}.
\newblock {\emph{\JournalTitle{Nano Lett.}}} \textbf{\bibinfo{volume}{24}}, \bibinfo{pages}{10577–10582} (\bibinfo{year}{2024}).
\newblock \doiprefix\url{10.1021/acs.nanolett.4c02765}.

\bibitem{Koshelev2023}
\bibinfo{author}{Koshelev, K.}, \bibinfo{author}{Tonkaev, P.} \& \bibinfo{author}{Kivshar, Y.}
\newblock \bibinfo{journal}{\bibinfo{title}{Nonlinear chiral metaphotonics: a perspective}}.
\newblock {\emph{\JournalTitle{Adv. Photonics}}} \textbf{\bibinfo{volume}{5}}, \bibinfo{pages}{064001} (\bibinfo{year}{2023}).
\newblock \doiprefix\url{10.1117/1.ap.5.6.064001}.

\bibitem{Wu_2024}
\bibinfo{author}{Wu, D.}, \bibinfo{author}{Ye, M.}, \bibinfo{author}{Chen, H.}, \bibinfo{author}{Xu, Y.} \& \bibinfo{author}{Duan, W.}
\newblock \bibinfo{journal}{\bibinfo{title}{Giant and controllable nonlinear magneto-optical effects in two-dimensional magnets}}.
\newblock {\emph{\JournalTitle{npj Comput. Mater.}}} \textbf{\bibinfo{volume}{10}}, \bibinfo{pages}{79} (\bibinfo{year}{2024}).
\newblock \doiprefix\url{10.1038/s41524-024-01266-x}.

\bibitem{Ono_2024}
\bibinfo{author}{Ono, A.}, \bibinfo{author}{Okumura, S.}, \bibinfo{author}{Imai, S.} \& \bibinfo{author}{Akagi, Y.}
\newblock \bibinfo{journal}{\bibinfo{title}{High harmonic generation from electrons moving in topological spin textures}}.
\newblock {\emph{\JournalTitle{Phys. Rev. B}}} \textbf{\bibinfo{volume}{110}}, \bibinfo{pages}{125111} (\bibinfo{year}{2024}).
\newblock \doiprefix\url{10.1103/PhysRevB.110.125111}.

\bibitem{Son2021}
\bibinfo{author}{Son, J.} \emph{et~al.}
\newblock \bibinfo{journal}{\bibinfo{title}{Air-stable and layer-dependent ferromagnetism in atomically thin van der \textrm{Waals} \textrm{CrPS}$_4$}}.
\newblock {\emph{\JournalTitle{ACS Nano}}} \textbf{\bibinfo{volume}{15}}, \bibinfo{pages}{16904–16912} (\bibinfo{year}{2021}).
\newblock \doiprefix\url{10.1021/acsnano.1c07860}.

\bibitem{Wu2022}
\bibinfo{author}{Wu, R.} \emph{et~al.}
\newblock \bibinfo{journal}{\bibinfo{title}{Magnetotransport study of van der \textrm{Waals} \textrm{CrPS$_4$/(Pt,Pd)} heterostructures: spin-flop transition and room-temperature anomalous hall effect}}.
\newblock {\emph{\JournalTitle{Phys. Rev. Appl.}}} \textbf{\bibinfo{volume}{17}}, \bibinfo{pages}{064038} (\bibinfo{year}{2022}).
\newblock \doiprefix\url{10.1103/PhysRevApplied.17.064038}.

\bibitem{Seo2024}
\bibinfo{author}{Seo, J.~Y.} \emph{et~al.}
\newblock \bibinfo{journal}{\bibinfo{title}{Probing the weak limit of magnetocrystalline anisotropy through a spin‒flop transition in the van der \textrm{Waals} antiferromagnet \textrm{CrPS}$_4$}}.
\newblock {\emph{\JournalTitle{NPG Asia Mater.}}} \textbf{\bibinfo{volume}{16}}, \bibinfo{pages}{39} (\bibinfo{year}{2024}).
\newblock \doiprefix\url{10.1038/s41427-024-00559-3}.

\bibitem{Budko2021}
\bibinfo{author}{Bud'ko, S.~L.}, \bibinfo{author}{Gati, E.}, \bibinfo{author}{Slade, T.~J.} \& \bibinfo{author}{Canfield, P.~C.}
\newblock \bibinfo{journal}{\bibinfo{title}{Magnetic order in the van der \textrm{Waals} antiferromagnet \textrm{CrPS}$_4$: \textrm{Anisotropic} $h\text{\ensuremath{-}}t$ phase diagrams and effects of pressure}}.
\newblock {\emph{\JournalTitle{Phys. Rev. B}}} \textbf{\bibinfo{volume}{103}}, \bibinfo{pages}{224407} (\bibinfo{year}{2021}).
\newblock \doiprefix\url{10.1103/PhysRevB.103.224407}.

\bibitem{Jungwirth2016}
\bibinfo{author}{Jungwirth, T.}, \bibinfo{author}{Marti, X.}, \bibinfo{author}{Wadley, P.} \& \bibinfo{author}{Wunderlich, J.}
\newblock \bibinfo{journal}{\bibinfo{title}{Antiferromagnetic spintronics}}.
\newblock {\emph{\JournalTitle{Nat. Nanotechnol.}}} \textbf{\bibinfo{volume}{11}}, \bibinfo{pages}{231–241} (\bibinfo{year}{2016}).
\newblock \doiprefix\url{10.1038/nnano.2016.18}.

\bibitem{Adachi1980}
\bibinfo{author}{Adachi, K.}, \bibinfo{author}{Achiwa, N.} \& \bibinfo{author}{Mekata, M.}
\newblock \bibinfo{journal}{\bibinfo{title}{Helical magnetic structure in \textrm{CsCuCl}$_3$}}.
\newblock {\emph{\JournalTitle{J. Phys. Soc. Jpn.}}} \textbf{\bibinfo{volume}{49}}, \bibinfo{pages}{545–553} (\bibinfo{year}{1980}).
\newblock \doiprefix\url{10.1143/jpsj.49.545}.

\bibitem{Sosnowska2002}
\bibinfo{author}{Sosnowska, I.}, \bibinfo{author}{Sch\:{a}fer, W.}, \bibinfo{author}{Kockelmann, W.}, \bibinfo{author}{Andersen, K.} \& \bibinfo{author}{Troyanchuk, I.}
\newblock \bibinfo{journal}{\bibinfo{title}{Crystal structure and spiral magnetic ordering of \textrm{BiFeO}$_3$ doped with manganese}}.
\newblock {\emph{\JournalTitle{Appl. Phys. A}}} \textbf{\bibinfo{volume}{74}}, \bibinfo{pages}{s1040–s1042} (\bibinfo{year}{2002}).
\newblock \doiprefix\url{10.1007/s003390201604}.

\bibitem{Arkenbout2006}
\bibinfo{author}{Arkenbout, A.~H.}, \bibinfo{author}{Palstra, T. T.~M.}, \bibinfo{author}{Siegrist, T.} \& \bibinfo{author}{Kimura, T.}
\newblock \bibinfo{journal}{\bibinfo{title}{Ferroelectricity in the cycloidal spiral magnetic phase of \textrm{MnWO}$_4$}}.
\newblock {\emph{\JournalTitle{Phys. Rev. B}}} \textbf{\bibinfo{volume}{74}}, \bibinfo{pages}{184431} (\bibinfo{year}{2006}).
\newblock \doiprefix\url{10.1103/PhysRevB.74.184431}.

\bibitem{Menzel2012}
\bibinfo{author}{Menzel, M.} \emph{et~al.}
\newblock \bibinfo{journal}{\bibinfo{title}{Information transfer by vector spin chirality in finite magnetic chains}}.
\newblock {\emph{\JournalTitle{Phys. Rev. Lett.}}} \textbf{\bibinfo{volume}{108}}, \bibinfo{pages}{197204} (\bibinfo{year}{2012}).
\newblock \doiprefix\url{10.1103/PhysRevLett.108.197204}.

\bibitem{KNB2005}
\bibinfo{author}{Katsura, H.}, \bibinfo{author}{Nagaosa, N.} \& \bibinfo{author}{Balatsky, A.~V.}
\newblock \bibinfo{journal}{\bibinfo{title}{Spin current and magnetoelectric effect in noncollinear magnets}}.
\newblock {\emph{\JournalTitle{Phys. Rev. Lett.}}} \textbf{\bibinfo{volume}{95}}, \bibinfo{pages}{057205} (\bibinfo{year}{2005}).
\newblock \doiprefix\url{10.1103/PhysRevLett.95.057205}.

\bibitem{Peng2020}
\bibinfo{author}{Peng, Y.} \emph{et~al.}
\newblock \bibinfo{journal}{\bibinfo{title}{Magnetic structure and metamagnetic transitions in the van der \textrm{Waals} antiferromagnet \textrm{CrPS}$_4$}}.
\newblock {\emph{\JournalTitle{Adv. Mater.}}} \textbf{\bibinfo{volume}{32}}, \bibinfo{pages}{202001200} (\bibinfo{year}{2020}).
\newblock \doiprefix\url{10.1002/adma.202001200}.

\bibitem{elk}
\bibinfo{title}{The elk code}.
\newblock \bibinfo{howpublished}{\url{https://elk.sourceforge.io/}}.
\newblock \bibinfo{note}{(accessed 2024-17-01)}.

\end{thebibliography}

\section*{Acknowledgements}

We thank S. Wu and Z. Wang for helpful discussions. This work was supported by the Research Council of Finland (349696, 352780, 352930, 353364, 360411, 358088, 359009, and 365686), the Flagship Program of the Research Council of Finland (PREIN; Quantum), the Villum Fonden (70202), the Independent Research Fund Denmark (2032-00351B), the EU H2020-MSCA-RISE-872049 (IPN-Bio), the Jane and Aatos Erkko foundation, and ERC (834742), NRF Korea (Grant No. 2021R1A2C3005905 and RS-2024-00466612). We acknowledge the computational resources provided by the Aalto Science-IT project. 

\section*{Author contributions}

Z.S. and Y.H. conceived the idea and designed the experiments. J.L.L. and Z.S. supervised the project. Y.H. built the experimental setup with the help of H.L., N.S. and Y.Z..  Y.H. prepared the sample and performed the experimental measurements with the help of M.T.. J.C.A. conducted the Raman measurement. T.V.C.A. conducted the theoretical calculation with the guidance of J.L.L. and A.O.F.. Y.H., Z.S., T.V.C.A. and J.L.L. analyzed all of the data and co-drafted the paper.  All authors reviewed the manuscript.

\section*{Competing interests}

There are no competing interests to declare.

\section*{Additional information}

Supplementary information is available.

\end{document}